\documentstyle[11pt,newpasp,twoside,epsfig]{article}
\def\edcomment#1{\iffalse\marginpar{\raggedright\sl#1\/}\else\relax\fi}
\reversemarginpar

\begin{document}
\title{Chemical and dynamical evolution in gas-rich dwarf galaxies}
\author{Simone Recchi, Francesca Matteucci}
\affil{Dipartimento di Astronomia, Universit\`a di Trieste, Via G.B. 
Tiepolo 11, 34131 Trieste, Italy}
\author{Annibale D'Ercole}
\affil{Osservatorio Astronomico di Bologna, Via Ranzani 1, 44127 Bologna, 
Italy}

\begin{abstract}
We study the effect of a single, instantaneous starburst in a gas-rich
dwarf galaxy on the dynamical and chemical evolution of its
interstellar medium. We consider the energetic input and the chemical
yields originating from SNeII, SNeIa and intermediate-mass stars. We
find that a galaxy resembling IZw18 develops a galactic wind carrying
out mostly the metal-rich gas. The various metals are lost
differentially and the metals produced by the SNeIa are lost more
efficiently than the others. As a consequence, we find larger
[$\alpha$/Fe] ratios for the gas inside the galaxy than for the gas
leaving the galaxy. Finally we find that a single burst occurring in
primordial gas (without pre-enrichment), gives chemical abundances and
dynamical structures in good agreement with what observed in IZw18
after $\sim$ 29 Myr from the beginning of star formation.

\end{abstract}

\section{Introduction}
Blue compact dwarf galaxies (BCD) are gas-rich systems experiencing an
intense star formation. These galaxies have very simple structures,
small sizes and are very metal poor. For these reasons, BCD are
excellent laboratories to investigate the effect of a starburst on the
chemical and dynamical evolution of the interstellar medium (ISM).

Previous dynamical and chemical studies of these galaxies have
suggested the existence of a `differential galactic wind', in the
sense that after a starburst event these objects would loose mostly
metals (ref. Mac Low \& Ferrara 1999; D'Ercole \& Brighenti 1999;
Pilyugin 1992, 1993; Marconi et al. 1994). However, in none of these
studies, detailed chemical and dynamical evolution was taken into
account at the same time. The aim of this paper is to include the
effects (both energetic and chemical) of type II and type Ia SNe in a
detailed dynamical model.

\section{Model description}

We consider a rotating gaseous component in hydrostatic isothermal
equilibrium with the gravitational and the centrifugal forces. The
potential well is given by the sum of a spherical, quasi-isothermal
dark halo and an oblate King profile. The resulting gas distribution
resembles that observed in IZw18 in a region $R\le$ 1 Kpc and $z\le$
730 pc, which we call `galactic region'. 

To describe the evolution of the ISM we solve a set of time-dependent,
hydrodynamical equations, with source terms describing the rate of
energy and mass return from the starburst. Mass is returned mostly by
SNeII and intermediate-mass stars (IMS), while the energy is injected
essentially by SNe. For the first time, here we take into account also
the contribution by SNeIa. These supernovae start to explode after 29
Myr, at the end of the SNII activity, occurring with the explosion of
stars with 8 M$_{\odot}$ (see Nomoto, Thielemann \& Yokoi 1984).

Following Bradamante et al. (1998), we suppose that SNeII convert only
3\% of their explosion energy into thermal energy of the ISM. SNeIa,
instead, do not suffer radiative losses because they explode in a
medium heated and diluted by the previous SNeII activity and
release all their energy into the ISM.

We solve an ancillary set of equations which keep track of the
evolution in space and time of some specific elements lost by stars,
namely H, He, C, N, O, Mg, Si, Fe. The production of these elements
are obtained following the nucleosynthetic prescriptions from various
authors: Woosley \& Weaver (1995) for the SNeII, Renzini \& Voli
(1981) for IMS and Nomoto et al. (1984) for SNeIa. For more details,
see Recchi et al. (2000).

The standard model, called M1, has a gaseous mass inside the galactic
region of $\sim 1.7\times 10^7\,{\rm M}_{\sun}$ and a mass of gas
turned into stars of $\sim 6\times 10^5\,{\rm M}_{\sun}$, in
reasonable agreement with the observations of IZw18. We run other two
models obtained by reducing the burst luminosity of a factor 0.6
(model M2) and by reducing the mass of gas of a factor 0.25 (model M3).
Moreover, we consider four nucleosynthetic options: we consider an
initial abundance of the ISM of $Z=0$ and $Z=0 .01\,{\rm Z}_{\odot}$
and two possible values for the mixing lenght parameter $\alpha_{\rm
RV}=0$ and $\alpha_{\rm RV}=1.5$. In the models with $\alpha_{\rm
RV}=1.5$ we can produce N in a primary way in IMS.

\section{Results}

In model M1 a classical bubble develops as a consequence of SNII
explosions (see Fig. 1). It expands faster along the $z$ direction,
where the ISM density gradient is steeper. The SNII wind stops before
the possible breakout, and the subsequent SNIa wind is not strong
enough to expand the cavity further. The size of the bubble thus does
not change for nearly 300 Myr, although the shape varies irregularly
because of the Kelvin-Helmholtz instabilities along the interface
between the hot cavity and the surrounding gas. After $\sim$ 340 Myr
the expanding ISM is diluted enough and the hot bubble finally breaks
out through a funnel. Most of the SNII ejecta remain locked into the
bubble wall inside the galaxy, while the SNIa elements, ejected later,
are easily channelled along the funnel. Iron is mostly produced by
SNeIa and, when the breakup occurs, most of it is lost. Thus the gas
[$\alpha$/Fe] ratio results lower outside the galaxy than inside (see
Fig. 2).

\begin{figure}
\plotone{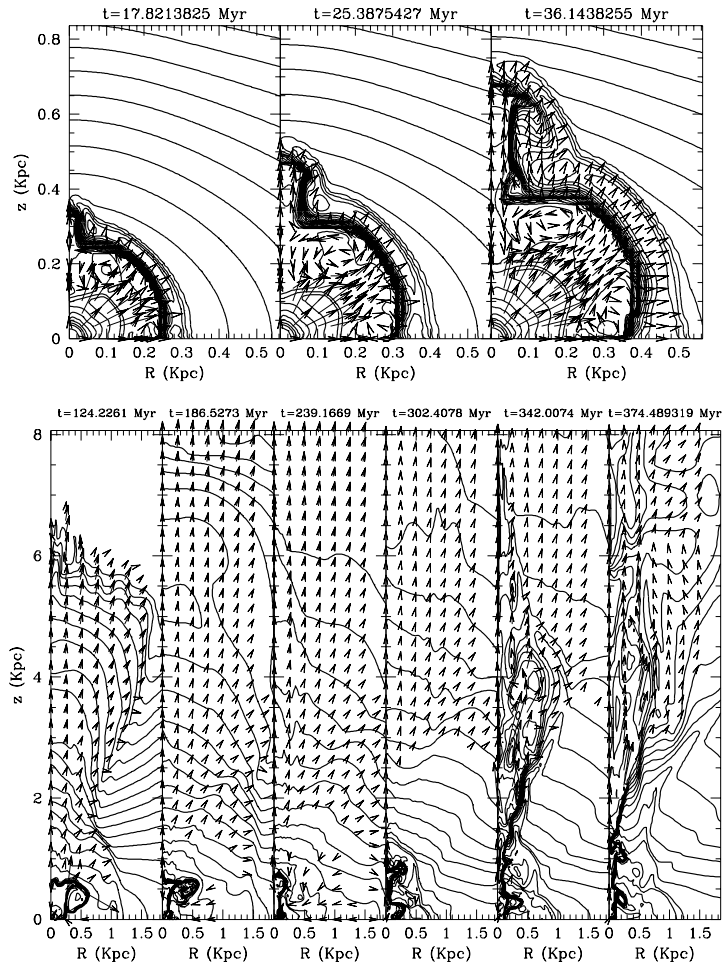}
\caption{Isodensity curves (in logarithmic scale) and velocity field for 
the model M1 at various burst ages.}
\end{figure}

\begin{figure}
\vspace{-6cm}
\plotone{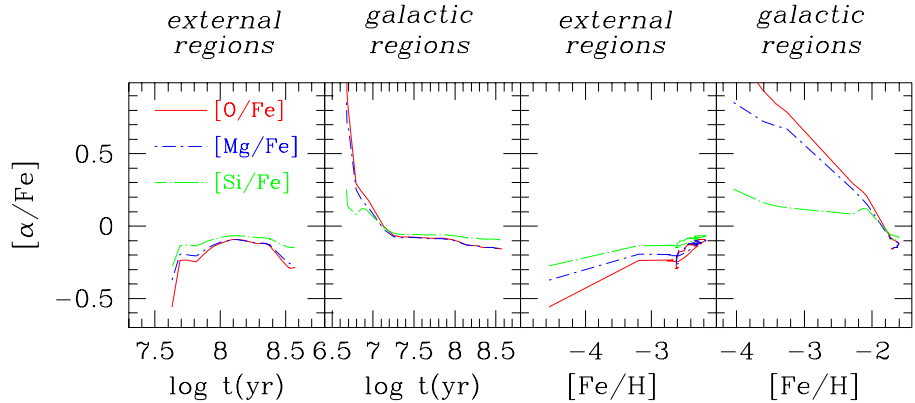}
\caption{Predicted [$\alpha$/Fe] vs. time and vs. [Fe/H] for both
expelled gas and ISM for the model M1}
\end{figure}

After $\sim$ 29 Myr [the burst in IZw18 is evaluated to be 15 - 27 Myr
old by Martin (1996)] the galactic abundances found in this model are
in good agreement with those observed in IZw18 once the
nucleosynthetic prescriptions with $Z=0$ and $\alpha_{\rm RV}=1.5$ are
assumed. At this time a substantial fraction of N is produced by IMS
in a primary way. An initial metallicity of $Z=0.01\,{\rm Z}_{\sun}$
(simulating a pre-enriched burst), worsens the agreement between data
and model results. Also the observed dimensions of the dynamical
structures are in reasonable agreement with our result after $\sim$ 29
Myr.

Models M2 and M3 have similar dynamical behaviours. However, due to
the different quantity of metals produced and gas mass lost, their
abundances are overestimated (M3) or underestimated (M2) compared to
IZw18. We also run a model similar to M1 but with a 100\% efficiency
of SNeII in heating the gas. In this case the galaxy results devoided
of gas 450 Myr after the burst, at variance with the substantial
amount of ISM in IZw18.

\end{document}